%% file: main.tex
\documentclass[11pt,a4paper]{article}
\pdfoutput=1
\usepackage{jinstpub}

\usepackage[ngerman,american]{babel}

\usepackage{microtype}

\usepackage{pgfplots}
\pgfplotsset{compat=1.16}

\usepackage{svg}

\usepackage{siunitx}
\DeclareSIUnit{\year}{year}
\DeclareSIUnit{\counts}{counts}
\DeclareSIUnit{\bin}{bin}
\DeclareSIUnit{\phiz}{\text{\ensuremath{{\Phi}_{0}}}}
\DeclareSIUnit{\promille}{\text{\textperthousand}}
\DeclareSIUnit{\ppm}{ppm}

\let\pgfimageWithoutPath\pgfimage 
\renewcommand{\pgfimage}[2][]{\pgfimageWithoutPath[#1]{gfx/#2}}

\title{High-resolution for IAXO: MMC-based X-ray Detectors}

\author{D.~Unger,}
\author{A.~Abeln,}
\author{C.~Enss,}
\author{A.~Fleischmann,}
\author{D.~Hengstler,}
\author{S.~Kempf}
\author{and L.~Gastaldo}

\affiliation{Kirchhoff Institute for Physics, Department of Physics and Astronomy, Heidelberg University\\Im Neuenheimer Feld 227, 69120 Heidelberg, Germany}

\emailAdd{daniel.unger@kip.uni-heidelberg.de}

\abstract{Axion helioscopes like the planned International Axion Observatory (IAXO) search for evidence of axions and axion-like particles (ALPs) from the Sun. A strong magnetic field is used to convert ALPs into photons via the generic ALP-photon coupling. To observe the resulting photons, X-ray detectors with low background and high efficiency are necessary. In addition, good energy resolution and low energy threshold would allow for investigating the ALP properties by studying the X-ray spectrum after its discovery. We propose to use low temperature metallic magnetic calorimeters (MMCs). Here we present the first detector system based on MMCs developed for IAXO and discuss the results of the characterization. The detector consists of a two-dimensional \num{64}-pixel array covering an active area of \SI{16}{\milli\metre\squared} with a fill factor of \SI{93}{\percent}. We achieve an average energy resolution of \SI{6.1}{\electronvolt} FWHM allowing for energy thresholds below \SI{100}{\electronvolt}. This detector is the first step towards a larger \SI{1}{\centi\metre\squared} array matching the IAXO X-ray optics. We determine the background rate for an unshielded detector system in the energy range between \SI{1}{\kilo\electronvolt} and \SI{10}{\kilo\electronvolt} to be \SI{3.2(1)e-4}{\per\kilo\electronvolt\per\centi\metre\squared\per\second} from events acquired over \num{30} days. In the future, active and passive shields will significantly reduce the background induced by cosmic muons and natural radioactivity. Our results demonstrate that MMCs are a promising technology  for helioscopes to discover and study ALPs.}


\arxivnumber{2010.15348}

\begin{document}
\maketitle
\input{txt/txt_introduction}
\input{txt/txt_maXs30}
\input{txt/txt_design}
\input{txt/txt_results}
\input{txt/txt_conclusion}
\input{txt/txt_acknowledgments}
\bibliographystyle{JHEP}
\bibliography{bibfile}
\end{document}

%% file: txt/txt_introduction.tex
\section{Introduction}\label{sec:introduction}

Axions are hypothetical particles, originally predicted by the Peccei–Quinn theory as a possible solution to the strong $\mathrm{CP}$ problem~\cite{Peccei_1977a,Peccei_1977b,Weinberg_1978,Wilczek_1978}. Via a fermion loop, they have a very weak coupling to photons. The coupling to photons as well as the axion mass are proportional to the inverse of the energy scale related to the spontaneous breaking of the Peccei–Quinn symmetry. Axions are therefore characterized by only one parameter. Particles having a similar two photon interaction, but with a not necessarily related mass and photon coupling, are called axion-like particles (ALPs) and are proposed in several theories beyond the standard model~\cite{Jaeckel_2010}. ALPs are of particular interest as they are also considered a well-motivated dark matter candidate~\cite{Preskill_1983,Abbott_1983,Arias_2012}. Their existence could also explain astrophysical observations like the $\gamma$-ray transparency of the Universe and stellar cooling anomalies~\cite{De_Angelis_2007,Mirizzi_2007,Isern_2008,Giannotti_2017}. Several experiments are looking for them using different methods differentiated by the investigated ALP source: light-shining-through-a-wall experiments are designed to produce and convert ALPs in laboratories~\cite{Baehre_2013}, haloscopes look for relic ALPs as part of the local dark matter halo~\cite{Du_2018}, whereas helioscopes search for ALPs generated in the Sun~\cite{Lazarus_1992,Ohta_2012,Anastassopoulos_2017}.

The Sun is potentially the strongest ALP source in our vicinity. The expected solar ALP flux can be described by two components, originating from ALP-photon and ALP-electron interactions respectively. Figure~\ref{fig:axion} shows the expected solar ALP spectrum on Earth, assuming an ALP-photon coupling $g_{\mathrm{a}\gamma}$ of $\SI{e-11}{\per\giga\electronvolt}$ and an ALP-electron coupling $g_\mathrm{ae}$ of $\num{e-13}$ as suggested by stellar cooling anomalies~\cite{Giannotti_2017}. ALPs from Primakoff conversion (orange, dashed) are generated by the interaction of black-body photons with virtual photons of the dense plasma in the interior of the Sun. The spectrum has a maximum at about \SI{3}{\kilo\electronvolt}, corresponding to the inner solar temperature. The spectrum from electron processes (blue, solid) has a smooth constituent with a maximum at about \SI{1}{\kilo\electronvolt} due to Bremsstrahlung and Compton scattering with outgoing ALPs. The resonances are due to ALP-recombination and ALP-deexcitation, which depend on the metal composition of the Sun~\cite{Redondo_2013}. The possibility to determine the relative intensity of the flux components will be important to identify the underlying ALP theory.

\begin{figure}[htb]
    \centering
    \input{gfx/gfx_1.pgf}
    \vspace{-0.1in}
    \caption{Expected solar ALP flux on Earth from electron processes (blue, solid) and Primakoff conversion (orange, dashed) assuming $g_{\mathrm{a}\gamma} = \SI{e-11}{\per\giga\electronvolt}$ and $g_\mathrm{ae} = \num{e-13}$~\cite{Jaeckel_2019b}.}
    \label{fig:axion}
\end{figure}

Helioscopes look for solar ALPs on Earth. In a helioscope, a long evacuated volume which is permeated by a strong magnetic field can be rotated and tilted to point towards the Sun for a large fraction of the day. The magnetic field is used to convert solar ALPs to more easily detectable X-rays via the generic ALP coupling to two photons~\cite{Sikivie_1983}. Three helioscopes have been built: the helioscope in Brookhaven~\cite{Lazarus_1992}, the Tokyo Axion Helioscope~\cite{Ohta_2012} and the CERN Axion Solar Telescope (CAST)~\cite{Anastassopoulos_2017}. So far, the most powerful helioscope is CAST which has set the current limit on $g_{\mathrm{a}\gamma}$ of \SI{6.6e-11}{\per\giga\electronvolt} for ALP masses $m_\mathrm{a}$ below \SI{0.02}{\electronvolt}~\cite{Anastassopoulos_2017}. The successor of CAST will be the International Axion Observatory (IAXO) with an expected sensitivity of a few \SI{e-12}{\per\giga\electronvolt} on  $g_{\mathrm{a}\gamma}$ for $m_\mathrm{a}$ up to \SI{0.01}{\electronvolt}~\cite{Armengaud_2014}. IAXO will have the potential to probe axion models in the \SI{1}{\milli\electronvolt} to \SI{1}{\electronvolt} mass range as well as an unexplored fraction of the ALP parameter space of particular interest where ALPs could be part of the cold dark matter and explain stellar cooling anomalies~\cite{Armengaud_2019}. This is technologically a very big step with respect to CAST and, therefore, the intermediate experiment BabyIAXO is currently under development to test major components like magnet, optics and X-ray detectors required for IAXO~\cite{Abeln_2020}. It will also be able to probe the existence of ALPs with $g_{\mathrm{a}\gamma}$ up to \SI{1.5e-11}{\per\giga\electronvolt} for $m_\mathrm{a}$ below \SI{0.02}{\electronvolt}.

Based on the expected very low ALP to photon conversion rate, ultra-low background X-ray detectors with a high efficiency up to \SI{10}{\kilo\electronvolt} and an active area of the order of \SI{0.2}{\centi\metre\squared} are required for IAXO. Gaseous time projection chambers (TPCs) equipped with Micromegas as used in CAST with a combination of active and passive shields achieve background rates below \SI{e-6}{\per\kilo\electronvolt\per\centi\metre\squared\per\second} and are considered as the baseline technology for BabyIAXO~\cite{Garza_2015}. However, different detector technologies with comparable efficiency and low background are essential to reduce systematic uncertainties in the interpretation of the data. At the same time, detectors with good energy resolution and low energy threshold are desired to study the solar ALP spectrum after discovery. In this case, the coupling strength of ALPs to photons and electrons as well as the underlying ALP model, could be identified by studying the spectrum in detail~\cite{Jaeckel_2019a}. Moreover, the ALP mass with $m_\mathrm{a}$ between \SI{3}{\milli\electronvolt} and \SI{100}{\milli\electronvolt} could be investigated from decoherence effects in ALP-photon oscillations~\cite{Dafni_2019}. Also information of the interior of the Sun like the metal composition and the solar magnetic field could be investigated~\cite{Jaeckel_2019b, O_Hare_2020}. Detectors based on low temperature metallic magnetic calorimeters (MMCs) feature good energy resolution and low energy threshold besides low intrinsic background and high quantum efficiency~\cite{Fleischmann_2005,Fleischmann_2009,Kempf_2018}. Therefore, MMCs are a perfect candidate to search for ALPs with helioscopes and in particular to study them beyond discovery.

We present the first MMC-based X-ray detector system developed for IAXO. In section~\ref{sec:maXs30}, we introduce the detector used for this system and describe the expected performance of the array. The design and the integration of the detector platform is depicted in section~\ref{sec:prototype}. In section~\ref{sec:results}, we show the results of the characterization, in particular the energy resolution and the background rate of the unshielded system. Finally, we review the achieved performance in section~\ref{sec:conclusion}.

%% file: txt/txt_maXs30.tex
\section{MaXs30 detector}\label{sec:maXs30}

Metallic magnetic calorimeters (MMCs) are operated at very low temperatures, usually below \SI{30}{\milli\kelvin}, and can reach remarkable energy resolution over a wide energy range~\cite{Fleischmann_2009}. They are used in various experiments due to their high resolving power $\frac{E}{\Delta E}$ up to \num{6000} and fast intrinsic response time, in the order of \SI{100}{\nano\second}, besides excellent linearity, high efficiency and low energy threshold~\cite{Pies_2012,Gastaldo_2017}. For example, a full width at half maximum (FWHM) energy resolution of \SI{1.6}{\electronvolt} was obtained for \SI{5.9}{\kilo\electronvolt} photons with a quantum efficiency of nearly \SI{100}{\percent}~\cite{Kempf_2018}. These properties, in combination with low intrinsic background, make MMC arrays a promising technology for helioscopes.

\begin{figure}[htb]
    \centering
    \footnotesize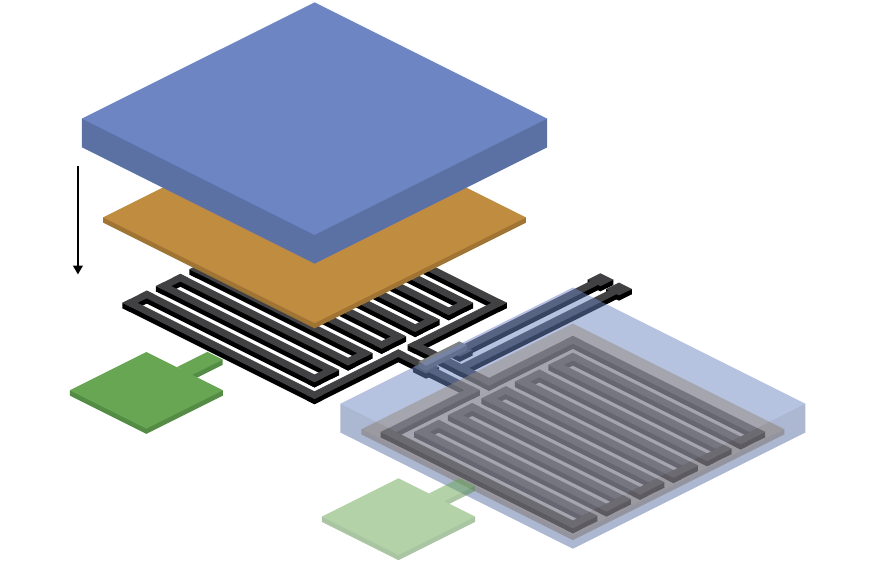
    \caption{Schematic drawing of a planar double meander geometry. Superconducting coils (black) connect two MMCs in parallel to a readout circuit. Each MMC consists of a particle absorber (blue) well thermally coupled to a paramagnetic sensor (orange) and is weakly coupled to a thermal bath (green).}
    \label{fig:mmc}
\end{figure}

The detection principle of MMCs is based on calorimetry. A typical design for MMCs is the so-called double meander geometry, shown in figure~\ref{fig:mmc}. This planar design allows for the operation of two pixels using one readout channel and the microfabrication of large and dense MMC arrays~\cite{Fleischmann_2005}. A single MMC pixel is composed of a particle absorber well thermally coupled to a paramagnetic temperature sensor sitting in a static magnetic field. When a particle interacts with the absorber, it deposits energy causing a small temperature increase. The temperature increase $\Delta T$ of absorber and sensor is approximately given by $\frac{E}{C}$, where $E$ is the energy deposited by the particle and $C$ is the total heat capacity of the MMC. The temperature increase of the sensor leads to a decrease of the magnetization $\Delta M$ given by $\frac{\partial M}{\partial T} \Delta T$ and creates a magnetic flux $\Delta \Phi$, proportional to $\Delta M$, in a superconducting pick-up coil directly underneath the sensor. The change of flux $\Delta \Phi$ is therefore proportional to $\frac{\partial M}{\partial T} \frac{E}{C}$ and thus proportional to the deposited energy of the particle. The flux change can be converted to a change of voltage using superconducting quantum interference devices (SQUIDs)~\cite{Clarke_2006}. A weak thermal link to a heat bath allows the MMC to slowly cool down to its operating temperature after the interaction of a particle.

In the case of the depicted double meander geometry, the superconducting pick-up coils underneath the two pixels are connected in parallel to the input coil of a {dc-SQUID} as indicated in figure~\ref{fig:mmc}. As a result, the two pick-up coils form a first order gradiometer which allows for distinguishing events in the two pixels by the polarity of the pulses and, in addition, this configuration reduces the effect of temperature fluctuations of the substrate on the output signal. The weak static magnetic field necessary to operate MMCs can be produced by a persistent current in the superconducting loop formed by the two meanders while the connection to the SQUID input coil is in its normal conducting state. The double meander geometry is also the basic design of the 32 channels of the maXs30 (micro-calorimeter array for X-ray spectroscopy) chip we chose for the first MMC-based detector system for BabyIAXO due to its relatively large active area with a high stopping power even above \SI{10}{\kilo\electronvolt} in combination with a good energy resolution~\cite{Hengstler_2015}. 

\begin{figure}[htbp]
    \centering
    {\footnotesize\color{white}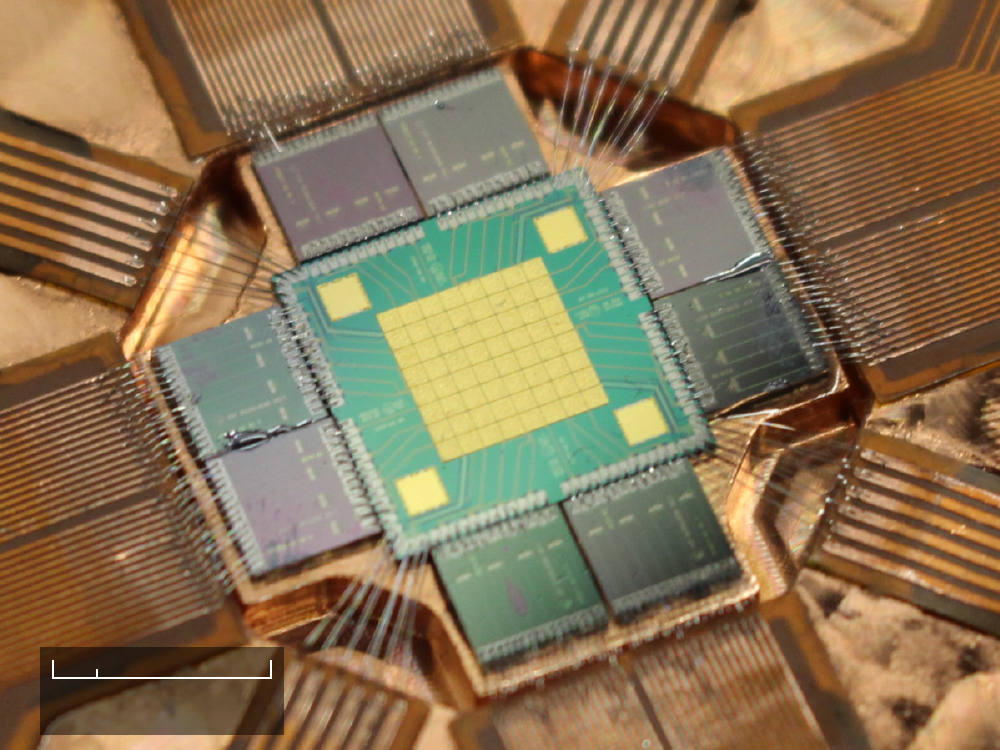}
    \caption{Photograph of the maXs30 chip glued on the copper platform together with eight first-stage SQUID chips. Electrical connections between the chips as well as to the polyimide circuit boards are provided by aluminum bonding wires which get superconducting.}
    \label{fig:maXs30}
\end{figure}

Figure~\ref{fig:maXs30} shows the maXs30 detector chip mounted on the newly developed copper platform together with eight first-stage SQUID chips, each hosting four SQUID channels, optimized for the readout of the MMCs. The detector and the SQUID chips were microfabricated in the cleanroom at the Kirchhoff Institute for Physics at Heidelberg University \cite{Kempf_2015}. The detector is a \num{64}-pixel two-dimensional MMC array, originally designed for experiments at the heavy ion storage ring ESR at the GSI and optimized for high-resolution X-ray spectroscopy up to \SI{30}{\kilo\electronvolt}~\cite{Hengstler_2015,Sikorsky_2020}. The maXs30 arrays are fabricated on three inch silicon wafers of about \SI{0.4}{\milli\metre} thickness. Each wafer contains \num{36} maXs30 chips with a size of $\SI{8}{\milli\metre} \times \SI{8}{\milli\metre}$ each. The absorbers are arranged in an eight by eight array with an area of \SI{16}{\milli\metre\squared}. Each absorber, made out of gold, has an area of $\SI{500}{\micro\metre} \times \SI{500}{\micro\metre}$ and a thickness of \SI{20}{\micro\metre} which guarantees a quantum efficiency higher than \SI{99}{\percent} for X-rays up to \SI{10}{\kilo\electronvolt}. For a small focal spot, the efficiency of the detector is limited by the fill factor of the absorbers and is given by \SI{93}{\percent}. The granularity of the array allows for a position sensitivity determined by the area of a single absorber. The temperature sensors with an area of $\SI{300}{\micro\metre} \times \SI{300}{\micro\metre}$ and a height of \SI{1.5}{\micro\metre} are made out of a dilute paramagnetic alloy of \SI{430}{\ppm} rare-earth metal erbium in the host material silver. The niobium meander-shaped pick-up coils have a line width of \SI{5}{\micro\metre}, a pitch of \SI{10}{\micro\metre} and a height of \SI{250}{\nano\metre}. The four double meanders at the corners of the array have a non-gradiometric design, obtained by reducing the area of one of the two sensors to $\SI{250}{\micro\metre} \times \SI{250}{\micro\metre}$. Due to this artificial asymmetry, the signal of these channels is sensitive to temperature fluctuations of the substrate and can be used to obtain the temperature of the detector chip.

The detector is optimized to operate at a temperature of \SI{20}{\milli\kelvin} with a persistent current of roughly \SI{70}{\milli\ampere}, which corresponds to an average magnetic field in the sensors of \SI{5}{\milli\tesla}. Under these conditions, the expected energy resolution $\Delta E_{\mathrm{FWHM}}$ is about $\SI{6}{\electronvolt}$. The voltage signal is completely characterized by an amplitude and the time constants of both, the exponential rise and decay. The amplitude is proportional to the energy deposited in the absorber during an event. The rise time is artificially limited by a thermal bottle neck between absorber and sensor which increases the intrinsic signal rise time to about \SI{10}{\micro\second}, else limited by the electron-spin coupling to \SI{100}{\nano\second}. Increasing the risetime is necessary to guarantee a position independent signal shape for particle interactions over the complete volume of the relatively large absorber. The decay time of about \SI{3}{\milli\second} is determined by the ratio of the total heat capacity of the MMC and the thermal conductance to the thermal bath, defined by the geometry of the gold thermal link. The pulse shape as well as the rise and decay time of different pixels vary slightly by a few percent due to inhomogeneities within the micro-structured layers and geometrical effects of the chip boundaries. Therefore, we perform the data analysis independently for each pixel.

Aluminum bonding wires, which are superconducting at the operating temperature, connect the double meander in parallel to input coils of {dc-SQUIDs} located on different chips. The MMCs generate signals in the SQUIDs of roughly \SI{10}{\milli\phiz\per\kilo\electronvolt} where $\si{\phiz} = \frac{h}{2 e}$ is the magnetic flux quantum. The signals from these first-stage SQUIDs are then amplified at \si{\milli\kelvin} temperatures using second-stage SQUID series-arrays~\cite{Mantegazzini_2020}. This two-stage SQUID readout scheme allows for reducing the noise contribution from the room temperature electronics. In this configuration, the first-stage SQUIDs are voltage biased via bias resistors fabricated on the same chips as the second-stage SQUIDs which reduces the power dissipation on the first-stage chips and, in turn, near the detector chip. The SQUID signal is linearized by room temperature electronics using a flux-locked-loop readout scheme~\cite{Drung_2006}.

%% file: gfx/gfx_2.pdf_tex
%% Creator: Inkscape 1.0 (4035a4fb49, 2020-05-01), www.inkscape.org
%% Accompanies image file 'gfx_364.pdf' (pdf, eps, ps)
\begingroup%
  \makeatletter%
  \providecommand\color[2][]{%
    \errmessage{(Inkscape) Color is used for the text in Inkscape, but the package 'color.sty' is not loaded}%
    \renewcommand\color[2][]{}%
  }%
  \providecommand\transparent[1]{%
    \errmessage{(Inkscape) Transparency is used (non-zero) for the text in Inkscape, but the package 'transparent.sty' is not loaded}%
    \renewcommand\transparent[1]{}%
  }%
  \providecommand\rotatebox[2]{#2}%
  \newcommand*\fsize{\dimexpr\f@size pt\relax}%
  \newcommand*\lineheight[1]{\fontsize{\fsize}{#1\fsize}\selectfont}%
  \ifx\svgwidth\undefined%
    \setlength{\unitlength}{252bp}%
    \ifx\svgscale\undefined%
      \relax%
    \else%
      \setlength{\unitlength}{\unitlength * \real{\svgscale}}%
    \fi%
  \else%
    \setlength{\unitlength}{\svgwidth}%
  \fi%
  \global\let\svgwidth\undefined%
  \global\let\svgscale\undefined%
  \makeatother%
  \begin{picture}(1,0.64285714)%
    \lineheight{1}%
    \setlength\tabcolsep{0pt}%
    \put(0,0){\includegraphics[width=\unitlength,page=1]{gfx/gfx_2.pdf}}%
    \put(0.735,0.37897442){\makebox(0,0)[lt]{\lineheight{1.25}\smash{\begin{tabular}[t]{l}$\mathrm{to~SQUID~input~coil}$\end{tabular}}}}%
    \put(0.61,0.56715178){\makebox(0,0)[lt]{\lineheight{1.25}\smash{\begin{tabular}[t]{l}$\mathrm{particle~absorber}$\end{tabular}}}}%
    \put(0.63,0.44171011){\makebox(0,0)[lt]{\lineheight{1.25}\smash{\begin{tabular}[t]{l}$\mathrm{paramagnetic~sensor}$\end{tabular}}}}%
    \put(-0.035,0.02772172){\makebox(0,0)[lt]{\lineheight{1.25}\smash{\begin{tabular}[t]{l}$\mathrm{superconducting~coil}$\end{tabular}}}}%
    \put(0,0){\includegraphics[width=\unitlength,page=2]{gfx/gfx_2.pdf}}%
    \put(-0.075,0.09045415){\makebox(0,0)[lt]{\lineheight{1.25}\smash{\begin{tabular}[t]{l}$\mathrm{thermal~bath}$\end{tabular}}}}%
    \put(0,0){\includegraphics[width=\unitlength,page=3]{gfx/gfx_2.pdf}}%
  \end{picture}%
\endgroup%

%% file: gfx/gfx_3.pdf_tex
%% Creator: Inkscape 1.0 (4035a4fb49, 2020-05-01), www.inkscape.org
%% Accompanies image file 'gfx_476.pdf' (pdf, eps, ps)
\begingroup%
  \makeatletter%
  \providecommand\color[2][]{%
    \errmessage{(Inkscape) Color is used for the text in Inkscape, but the package 'color.sty' is not loaded}%
    \renewcommand\color[2][]{}%
  }%
  \providecommand\transparent[1]{%
    \errmessage{(Inkscape) Transparency is used (non-zero) for the text in Inkscape, but the package 'transparent.sty' is not loaded}%
    \renewcommand\transparent[1]{}%
  }%
  \providecommand\rotatebox[2]{#2}%
  \newcommand*\fsize{\dimexpr\f@size pt\relax}%
  \newcommand*\lineheight[1]{\fontsize{\fsize}{#1\fsize}\selectfont}%
  \ifx\svgwidth\undefined%
    \setlength{\unitlength}{288bp}%
    \ifx\svgscale\undefined%
      \relax%
    \else%
      \setlength{\unitlength}{\unitlength * \real{\svgscale}}%
    \fi%
  \else%
    \setlength{\unitlength}{\svgwidth}%
  \fi%
  \global\let\svgwidth\undefined%
  \global\let\svgscale\undefined%
  \makeatother%
  \begin{picture}(1,0.75)%
    \lineheight{1}%
    \setlength\tabcolsep{0pt}%
    \put(0,0){\includegraphics[width=\unitlength,page=1]{gfx/gfx_3.pdf}}%
    \put(0.13000655,0.03322385){\makebox(0,0)[lt]{\lineheight{1.25}\smash{\begin{tabular}[t]{l}$\SI{5}{\milli\metre}$\end{tabular}}}}%
    \put(0,0){\includegraphics[width=\unitlength,page=2]{gfx/gfx_3.pdf}}%
  \end{picture}%
\endgroup%

%% file: txt/txt_design.tex
\section{System design}\label{sec:prototype}

The detector system developed in this work was designed to be suitable for the installation as a focal plane detector in the BabyIAXO helioscope. The detector platform is dimensioned to host MMC-based detector chips with a size up to $\SI{24}{\milli\metre} \times \SI{24}{\milli\metre}$. This gives flexibility to choose a detector geometry optimized for the focal plane defined by the X-ray optics \cite{Abeln_2020}. In addition, we have chosen a simple and modular design which allows to easily improve and exchange individual components as well as to add active and passive shields in the future. For the fabrication of the setup, we selected high purity materials to reduce the presence of radioactive contamination near the detector.

\begin{figure}[htbp]
    \centering
    \footnotesize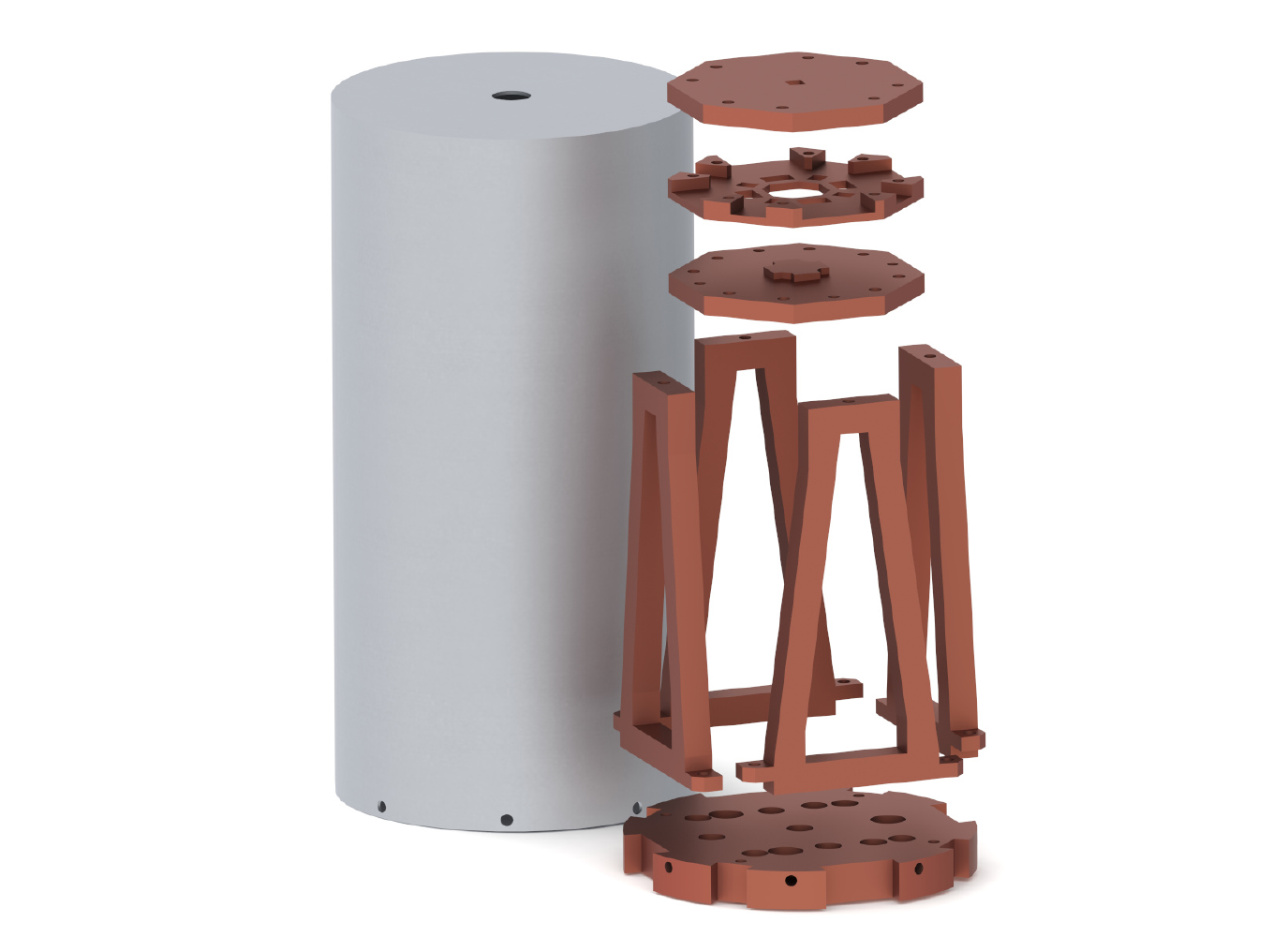
    \caption{Rendered image of the detector platform consisting of several copper parts and a niobium cover. The main part hosting the detector is the detector module assembled out of three octagonal copper parts.}
    \label{fig:solidworks}
\end{figure}

Figure~\ref{fig:solidworks} shows a rendered image of the designed metal components of the platform consisting of several copper parts and a niobium cover acting as a superconducting shield while cooled down below \SI{9.3}{\kelvin}. All copper parts are made out of oxygen-free high thermal conductivity (OFHC) copper\footnote{Allmeson GmbH, Ottostraße 9-11, 63150 Heusenstamm, Germany} with a purity of at least \SI{99.99}{\percent} and have been annealed after manufacturing to achieve a better heat conductivity at low temperatures and thus a lower operating temperature of the detector. We have chosen niobium\footnote{Haines \& Maassen Metallhandelsgesellschaft mbH, Pützchens Chaussee 60, 53227 Bonn, Germany} with a purity of at least \SI{99.9}{\percent} for the material of the superconducting shield due to its very high critical temperature. The detector and SQUID chips were glued onto the dedicated copper parts with a bicomponent epoxy\footnote{Huntsman International LLC, 10003 Woodloch Forest Drive, The Woodlands, Texas 77380, United States}. This type of glue is also applied in the Cryogenic Underground Observatory for Rare Events (CUORE) experiment and was tested to have low radioactive contamination~\cite{Alduino_2016}. The electrical connections from the detector module to the amplifier module are realized by flexible polyimide circuit boards\footnote{Multi Leiterplatten GmbH, Brunnthaler Straße 2, 85649 Brunnthal, Germany} with low radioactivity. To further reduce potential radioactivity, the circuit boards were manufactured neither with a stiffer layer nor a surface finish.

\begin{figure}[htbp]
    \centering
    \includegraphics[width=4in]{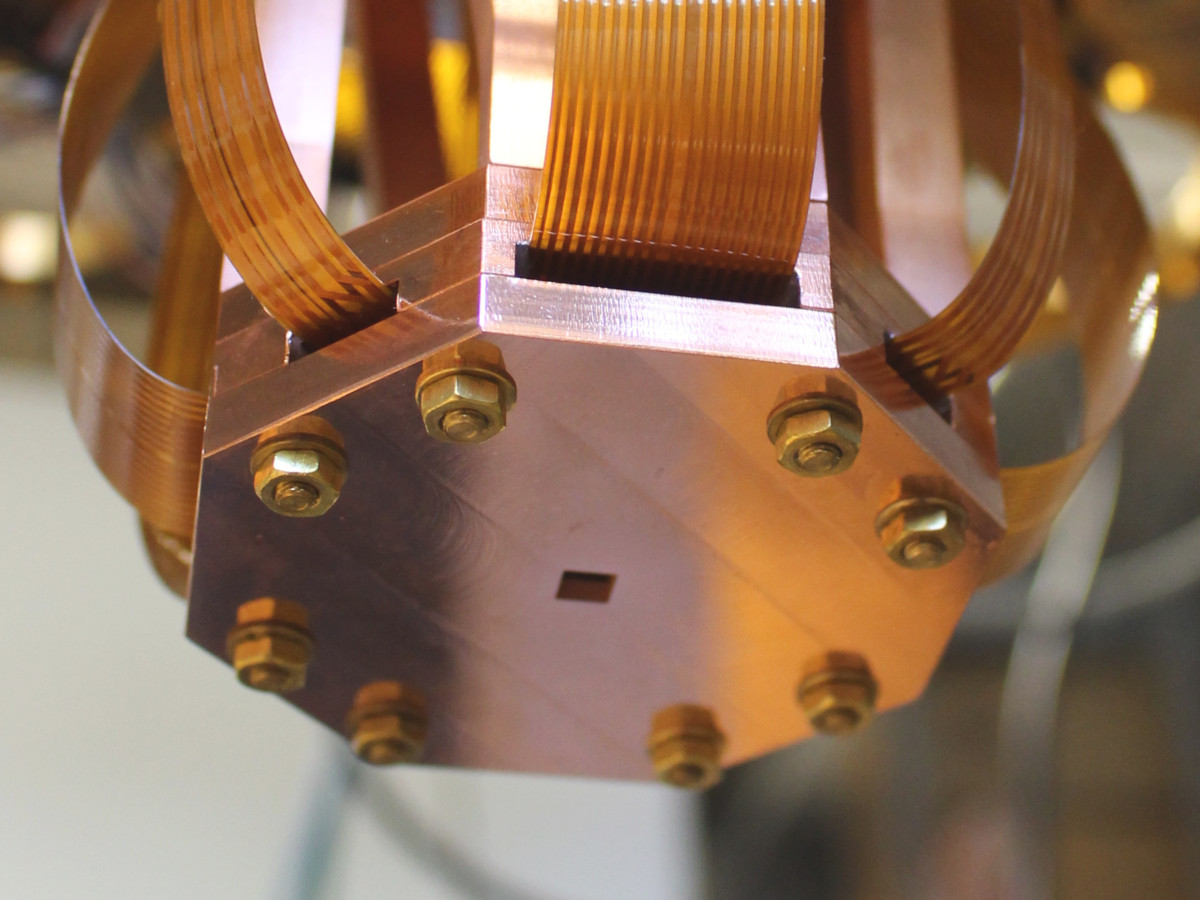}
    \caption{Octagonal detector module consisting of three copper parts. Polyimide circuit boards are used to connect the detector module to a SQUID amplifier module.}
    \label{fig:sandwich}
\end{figure}

The main component of the system is the detector module which consists of three copper parts. On the lower copper part of the detector module shown in figure~\ref{fig:solidworks}, the detector and eight first-stage SQUID chips are glued on a raised area in the center. Eight polyimide circuit boards are glued on the second copper part which has a hole in the center matching the raised area of the first part. Both parts are afterwards screwed together. The chips and circuit boards are then electrically connected with aluminum bonding wires, shown in figure~\ref{fig:maXs30}. The third part of the detector module is a collimator which is fixed on top of the other two parts. The complete detector module is shown in figure~\ref{fig:sandwich}.

\begin{figure}[htbp]
    \centering
    \includegraphics[width=3in]{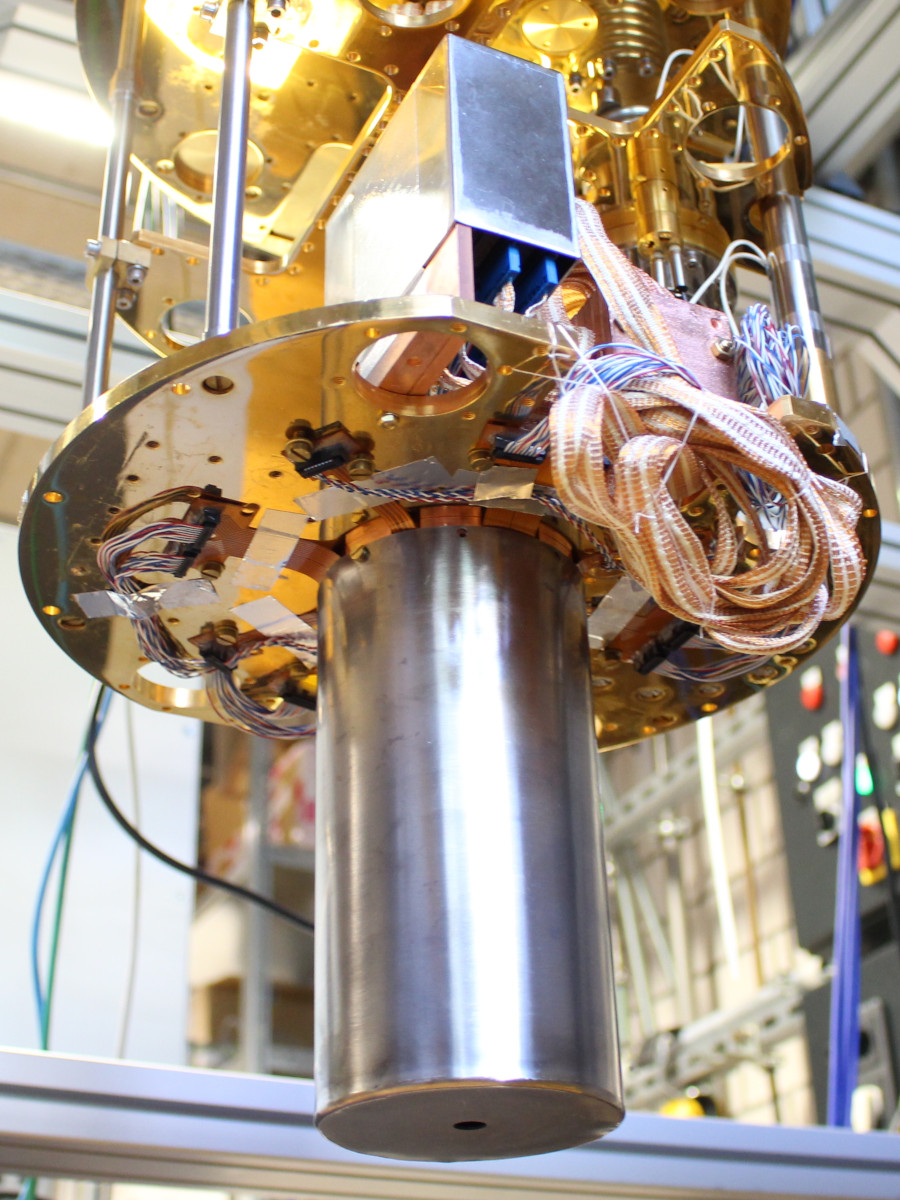}
    \caption{Complete system, mounted inside a cryostat and covered by the cylindrical superconducting niobium shield. The amplifier module is placed inside a rectangular cryoperm shield and is mounted on top of the mixing chamber plate. Ribbon cables connect the installation to room temperature electronics.}
    \label{fig:setup}
\end{figure}

The octagonal detector module with a distance between parallel sides of \SI{6}{\centi\metre} and a height of \SI{1.5}{\centi\metre} is mounted with four triangle shaped copper support structures to a copper adapter plate which can be screwed to the mixing chamber plate of a cryostat. The triangle structure prevents vibrations and rotations of the detector module whereas the adapter plate is designed to match the mounting holes of one of our dilution refrigerators\footnote{Bluefors Oy, Arinatie 10, 00370 Helsinki, Finland}. We use a tiny amount of vacuum grease between the copper parts except for the detector module to increase the thermal conductance. The niobium cover, acting as a superconducting shield, is screwed to the adapter plate to protect the SQUIDs and MMCs from magnetic field fluctuations. The complete system mounted inside a dilution refrigerator is shown in figure~\ref{fig:setup}. The niobium shield has a height of \SI{18}{\centi\metre} and a diameter of \SI{9}{\centi\metre}. Holes in the copper collimator and the niobium shielding allow the usage of external X-ray sources for characterization. For the discussed measurements, the source is positioned outside the cryostat at room temperature in front of an X-ray window. Other X-rays windows were also present in each of the thermal shields.

The wide polyimide circuit boards for the SQUID operation have standardized 16-pin connectors at the end which are connected to the SQUID amplifier module with cables as shown in figure~\ref{fig:setup}. The amplifier module as well as the narrow polyimide circuit boards are connected to ribbon cables. These cables, each with \num{30} wires made of copper with \SI{2}{\percent} nickel and having a diameter of \SI{200}{\micro\metre} each and a length of about \SI{2}{\metre}, are thermalized at each temperature stage of the dilution refrigerator and are connected at room temperature to 24-pin connectors\footnote{LEMO S.A., Chemin de Champs-Courbes 28, 1024 Ecublens, Switzerland} positioned in a vacuum tight aluminum box mounted on top of the cryostat~\cite{Mantegazzini_2020}. There, SQUID electronics\footnote{Magnicon GmbH, Barkhausenweg 11, 22339 Hamburg, Germany} for the flux-locked-loop readout are connected. Two ADCs\footnote{Struck Innovative Systeme GmbH, Harksheider Straße 102A, 22399 Hamburg, Germany} with 16 channels each are used for data acquisition. The signals from the 32 SQUID channels are acquired by triggering each channel individually and are passed from the ADCs to a dedicated PC via Ethernet.

%% file: gfx/gfx_4.pdf_tex
%% Creator: Inkscape 1.0 (4035a4fb49, 2020-05-01), www.inkscape.org
%% Accompanies image file 'gfx_926.pdf' (pdf, eps, ps)
\begingroup%
  \makeatletter%
  \providecommand\color[2][]{%
    \errmessage{(Inkscape) Color is used for the text in Inkscape, but the package 'color.sty' is not loaded}%
    \renewcommand\color[2][]{}%
  }%
  \providecommand\transparent[1]{%
    \errmessage{(Inkscape) Transparency is used (non-zero) for the text in Inkscape, but the package 'transparent.sty' is not loaded}%
    \renewcommand\transparent[1]{}%
  }%
  \providecommand\rotatebox[2]{#2}%
  \newcommand*\fsize{\dimexpr\f@size pt\relax}%
  \newcommand*\lineheight[1]{\fontsize{\fsize}{#1\fsize}\selectfont}%
  \ifx\svgwidth\undefined%
    \setlength{\unitlength}{396bp}%
    \ifx\svgscale\undefined%
      \relax%
    \else%
      \setlength{\unitlength}{\unitlength * \real{\svgscale}}%
    \fi%
  \else%
    \setlength{\unitlength}{\svgwidth}%
  \fi%
  \global\let\svgwidth\undefined%
  \global\let\svgscale\undefined%
  \makeatother%
  \begin{picture}(1,0.72727273)%
    \lineheight{1}%
    \setlength\tabcolsep{0pt}%
    \put(0,0){\includegraphics[width=\unitlength,page=1]{gfx/gfx_4.pdf}}%
    \put(0.00060388,0.39363676){\makebox(0,0)[lt]{\lineheight{1.25}\smash{\begin{tabular}[t]{l}$\mathrm{superconducting~shield}$\end{tabular}}}}%
    \put(0,0){\includegraphics[width=\unitlength,page=2]{gfx/gfx_4.pdf}}%
    \put(0.80378569,0.57443855){\makebox(0,0)[lt]{\lineheight{1.25}\smash{\begin{tabular}[t]{l}$\mathrm{detector~module}$\end{tabular}}}}%
    \put(0.80378762,0.28454836){\makebox(0,0)[lt]{\lineheight{1.25}\smash{\begin{tabular}[t]{l}$\mathrm{support~structures}$\end{tabular}}}}%
    \put(0.80378569,0.07545556){\makebox(0,0)[lt]{\lineheight{1.25}\smash{\begin{tabular}[t]{l}$\mathrm{adapter~plate}$\end{tabular}}}}%
    \put(0,0){\includegraphics[width=\unitlength,page=3]{gfx/gfx_4.pdf}}%
    \put(0.16421133,0.23120117){\color[rgb]{0.15686275,0.04313725,0.04313725}\makebox(0,0)[lt]{\lineheight{1.25}\smash{\begin{tabular}[t]{l}$\SI{5}{\centi\metre}$\end{tabular}}}}%
    \put(0,0){\includegraphics[width=\unitlength,page=4]{gfx/gfx_4.pdf}}%
  \end{picture}%
\endgroup%

%% file: txt/txt_results.tex
\section{Results}\label{sec:results}

We have characterized the detector at different temperatures and with different persistent currents to operate the MMCs at different magnetic fields. The used dilution refrigerator reaches a temperature below \SI{7}{\milli\kelvin} at the mixing chamber plate. Comparing the amplitude of the acquired signals with amplitudes obtained by calculations based on the thermodynamical properties of the MMCs, we find that the base temperature of the cryostat corresponds to a detector temperature of \SI{15(1)}{\milli\kelvin}. The temperature difference originates from the heat produced by the first-stage SQUIDs near the detector.

\subsection{Detector performance}

For the calibration of the detector system we used an $^{55}\mathrm{Fe}$~source\footnote{Eckert \& Ziegler, Robert-Rössle-Straße 10, 13125 Berlin, Germany}\addtocounter{footnote}{-1}\addtocounter{Hfootnote}{-1} as well as an $^{241}\mathrm{Am}$~source{\footnotemark}  for the characterization at higher energies. Both are closed sources, such that only X-rays can leave the housing. The radioactive sources were periodically positioned in front of the outer X-rays window of the cryostat. The response of the detector upon the absorption of $\mathrm{K}_\alpha$ photons at about \SI{5.9}{\kilo\electronvolt} from the $^{55}\mathrm{Fe}$ source is used to characterize the performance of the detector. To obtain the characteristic pulse shape, a few thousand pulses of this energy were averaged for each pixel. The averaged pulse is then scaled and fit to all acquired signals from the same pixel. This allows for the derivation of several parameters, in particular the signal amplitude and variables related to the pulse shape. Since the amplitude of the signal depends on the detector temperature, for each acquired trace we also record the output voltage of non-gradiometric detector channels which provide information on the chip temperature at the time the signal has been triggered. As a result, we can study the correlation between the temperature information and the amplitude of the signal and thus can correct for temperature fluctuations of the detector chip. In fact, slow temperature variations of the chip of the order of \SI{10}{\micro\kelvin} which induce variations on the signal amplitude of the order of \SI{0.5}{\percent} would else decrease the resolving power. To calibrate the signal amplitudes, we use the known energy of the $\mathrm{K}_\alpha$ lines as well as the $\mathrm{K}_\beta$ lines at about \SI{6.5}{\kilo\electronvolt} and adapt a quadratic fit to match the temperature corrected amplitude to the corresponding energy for each channel. We get a nonlinearity of roughly \SI{0.1}{\percent} at \SI{6}{\kilo\electronvolt}.

\begin{figure}[htb]
    \centering
    \input{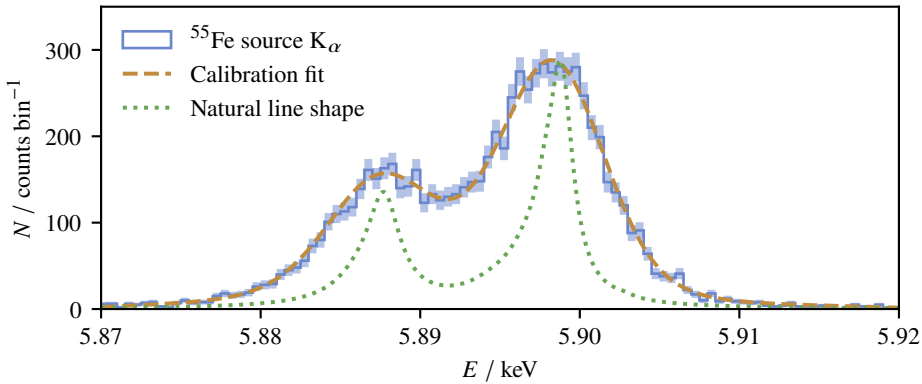}
    \vspace{-0.1in}
    \caption{Histogram obtained in one of the calibration measurements with an $^{55}\mathrm{Fe}$ source (blue, solid) for a single pixel. The Poisson uncertainty drawn on the bins is given by $\sqrt{N}$ where $N$ is the number of counts in the respective bin. The histogram has 100 bins with a bin width of \SI{0.5}{\electronvolt}. The FWHM energy resolution of \SI{6.4(2)}{\electronvolt} is determined by a calibration fit (orange, dashed). The natural line shape (green, dotted) shown for comparison is scaled to the maximum of the calibration fit.}
    \label{fig:iron}
\end{figure}

\begin{figure}[htb]
    \centering
    \input{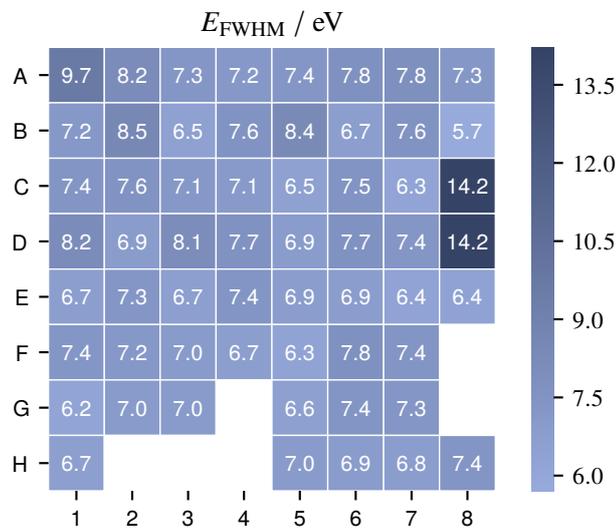}
    \vspace{-0.1in}
    \caption{Distribution of the FWHM energy resolution given by the Gaussian detector response evaluated at \SI{5.9}{\kilo\electronvolt} for the pixels operated during the first characterization run. The uncertainty is about \SI{0.2}{\electronvolt}. The average FWHM energy resolution is \SI{7.2}{\electronvolt} excluding the channel C8/D8 with a significantly higher noise.}
    \label{fig:eres}
\end{figure}

As an example, the histogram of the $\mathrm{K}_\alpha$ multiplett from the $^{55}\mathrm{Fe}$ source acquired for a single pixel during multiple calibration measurements is shown in figure \ref{fig:iron}. We fit the convolution of the intrinsic shape of the $\mathrm{K}_\alpha$ lines based on~\cite{Hoelzer_1997} and a Gaussian detector response with variable width to the histogram. The obtained Gaussian full width at half maximum (FWHM) of \SI{6.4(2)}{\electronvolt} represents the energy resolution of the MMC. Figure~\ref{fig:eres} shows, over a map representing the 64 pixels of the maXs30 chip, the FWHM energy resolution for the channels which have been operated during the discussed characterization run. Three of 32-channels could not be operated: two of them had a missing electrical connection at the SQUID amplifier level while for the third one the first-stage {dc-SQUID} had a visible damage. The three channels can be repaired for future experiments. Excluding the channel C8/D8 with a significantly higher noise, we obtained an average FWHM energy resolution of \SI{7.2}{\electronvolt} in this run. An evaluation of the energy resolution at \SI{0}{\electronvolt} via a baseline analysis yielded across \num{27} channels an average baseline energy resolution of \SI{6.1}{\electronvolt} FWHM which is in very good agreement with the expected \SI{6}{\electronvolt}. The baseline energy resolution was analyzed at a mixing chamber temperature of \SI{12}{\milli\kelvin} which corresponds to a slightly increased detector temperature of \SI{17(1)}{\milli\kelvin}. The very good energy resolution allows us to define very low trigger thresholds below \SI{100}{\electronvolt}.

\subsection{Background rate}

\begin{figure}[htb]
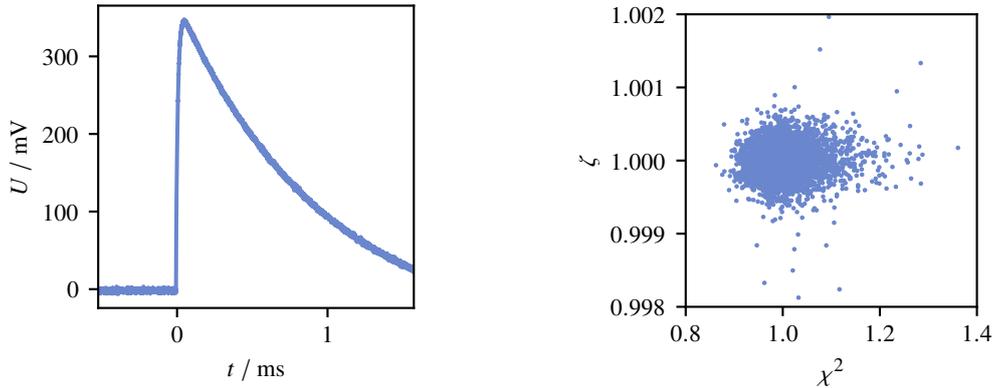

    %\centering
    \begin{minipage}{.5\textwidth}\centering\input{gfx/gfx_10.pgf}\end{minipage}\hfill%
    \begin{minipage}{.5\textwidth}\centering\input{gfx/gfx_11.pgf}\end{minipage}
    \vspace{-0.1in}
    \caption{Single, AC filtered pulse at \SI{5.9}{\kilo\electronvolt} from the $^{55}\mathrm{Fe}$ source (left). Pulses from the same pixel have a characteristic shape with the same rise and decay time constants. The $\chi^2$ - $\zeta$ plane, representing variations from the characteristic shape has an elliptical form (right). This contribution can be used to define a cut, identifying pulses with even tiny distortions from the characteristic shape.}
    \label{fig:ellipse}
\end{figure}

To determine the background of the detector it is important to distinguish events that are related to actual X-ray absorption within the aborber from other sources. As already mentioned, particles depositing their full energy in the MMC absorber lead to signals having a characteristic rise and decay time which is independent on the deposited energy for the energy range of interest. The typical shape of a single AC filtered \SI{5.9}{\kilo\electronvolt} pulse is shown in figure~\ref{fig:ellipse} on the left hand side. Charged particles passing through the absorber will have a chance to release some energy via ionization in the sensor or in the substrate close to the sensor or in both. This leads to modifications of the signal shape which can be recognized through pulse shape analysis. Furthermore, such particles can produce possible coincidence events in neighboring pixels. 

We use the two parameters $\chi^2$ and $\zeta$ to select events, whose full energy was deposited within the absorber. Both parameters are related to the shape of single traces compared to the characteristic pulse shape. Here, $\chi^2$ is given by the sum over each sample of the quadratic difference between a pulse and the fitted average pulse divided by the number of samples. This is also well known as the reduced $\chi^2$ from $\chi^2$ tests. The parameter $\zeta$ is based on a matched filter. To calculate $\zeta$ for a given pulse, two cross-correlations are performed: the pulse with the average pulse as well as the average pulse with itself . The parameter $\zeta$ is then an amplitude ratio given by the ratio of the two maxima divided by the ratio of the two integrals over the convolution \cite{Goeggelmann_2021}. On the right hand side of figure \ref{fig:ellipse} the $\chi^2$ - $\zeta$ plot corresponding to the calibration data from the histogram in figure \ref{fig:iron} is shown. Based on the analysis of the calibration measurements with external sources, we define an area with the shape of an ellipse in the $\chi^2$ - $\zeta$ plane. The semi-axes of the ellipse are determined by Gaussian fits, evaluating the form of the $\chi^2$ and $\zeta$ distributions for each pixel. For the discussed data analysis we set it to multiples of the Gaussian widths so that roughly \SI{1}{\percent} of the calibration events are located outside of this region and are rejected. We apply the same cut also to background measurements performed over a period of several days between two calibration runs. This ellipse cut yet shows an energy dependent efficiency for events having an energy lower than \SI{500}{\electronvolt} leading to a loss of rejection efficiency. For the background analysis we will consider the energy range between \SI{1}{\kilo\electronvolt} and \SI{10}{\kilo\electronvolt}, which is the range most interesting for IAXO. Improved algorithms for the data analysis are at present under development, promising a reliable pulse shape cut also at energies below \SI{500}{\electronvolt} with an efficiency loss less than \SI{1}{\percent} \cite{Barth_2020}. Typical signals which are removed with this approach are triggered noise, pile-up events and events related to massive particles releasing some of their energy in the sensor or in the substrate.

Very often triggered noise traces occur as burst of signals. To remove those traces during the background measurement we removed all recorded traces that where acquired within one minute if a threshold of 30 events per minute was exceeded in one of the two ADCs. Furthermore, one additional minute was removed before and after such a burst. The constraint was set such that signals induced from communication devices like mobile phones, creating many signals per minute can be easily detected while random background coincidences are very likely never affected. This cut reduces the effective measurement time by only \SI{5}{\percent} while we reduce the number of events by nearly two orders of magnitude. Noise traces which are not part of a burst and thus not affected by the burst cut usually contain instantaneous jumps, which by design cannot be produced from MMCs, and are easily identified with the ellipse cut. To remove fluorescence and particle showers that could for example be generated by muons interacting in the surrounding materials, we also removed all signals that were simultaneously triggered within \SI{1}{\micro\second} by more than one channel.

\begin{figure}[htb]
    \centering
    \input{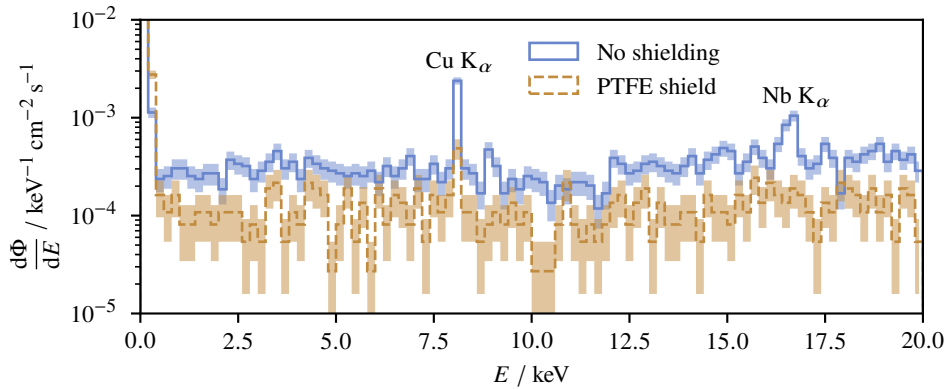}
    \vspace{-0.1in}
    \caption{Comparison between the background flux after pulse shape analysis, burst cut and coincidence cut without shielding (blue, solid) and with an additional PTFE shield (orange, dashed). The histogram has 100 bins with a bin width of \SI{200}{\electronvolt}. The Poisson uncertainty is given by $\sqrt{N}$.}
    \label{fig:background}
\end{figure}

During the first background analysis, we acquired about one month of raw background data with multiple calibration measurements in between to verify the stable operation of the system. Figure~\ref{fig:background} shows the background spectrum for the unshielded detector obtained after applying the described cuts (blue, solid). Between \SI{1}{\kilo\electronvolt} and \SI{10}{\kilo\electronvolt} the background rate is \SI{3.2(1)e-4}{\per\kilo\electronvolt\per\centi\metre\squared\per\second}. One can clearly identify copper $\mathrm{K}_\alpha$ lines at \SI{8.0}{\kilo\electronvolt} and the niobium $\mathrm{K}_\alpha$ lines at \SI{16.6}{\kilo\electronvolt}. Both fluorescence lines are potentially originating from the interactions of muons or with small probability by natural radioactivity. Minimal radioactive contamination of the materials used for the detector system might also contribute to the fluorescence in copper and niobium as well as to the energy-independent background spectrum.

At the Canfranc Underground Laboratory the intrinsic radioactive contamination of samples from the used copper, niobium and polyimide parts were analyzed with the help of low-background germanium detectors~\cite{Aznar_2013}. For the copper sample only upper activity limits were given. In the \SI{490}{\gram} niobium shield, $^{94}\mathrm{Nb}$ with an activity of \SI{33(3)}{\milli\becquerel\per\kilo\gram} was detected. From the $^{232}\mathrm{Th}$ chain, an activity of \SI{8.7(24)}{\milli\becquerel\per\kilo\gram} from $^{228}\mathrm{Ac}$ and \SI{8.8(23)}{\milli\becquerel\per\kilo\gram} from $^{228}\mathrm{Th}$ was found, hinting at a secular equilibrium. For the polyimide circuit boards, activities of \SI{30(11)}{\milli\becquerel\per\kilo\gram} and \SI{40(12)}{\milli\becquerel\per\kilo\gram} were found from $^{212}\mathrm{Pb}$ originating from the $^{232}\mathrm{Th}$ chain and $^{226}\mathrm{Ra}$ from the $^{238}\mathrm{U}$ chain respectively. For the system described in this work, polyimide circuit boards with a total mass of roughly \SI{11}{\gram} are used. A detailed simulation is required to determine the effect of the material contamination on the acquired background spectrum which is out of the scope of this publication. Nevertheless, we are at present designing a new superconducting shield based on copper which is plated with a superconducting film like tin~\cite{Mantegazzini_2020}.

Some of the detected fluorescence events have a relatively low energy and could be screened by materials with low atomic number placed between the collimator and the detector. In the second characterization run we studied the effect of a polytetrafluoroethylene (PTFE) piece with a diameter of \SI{43}{\milli\metre} and a thickness of \SI{4.5}{\milli\metre} on the background spectrum. The PTFE piece has a large squared $\SI{1}{\centi\metre} \times \SI{1}{\centi\metre}$ inner hole, since it was designed for a new, larger MMC array for the BabyIAXO experiment which is still in production. We were able to repair two of the three broken channels by replacing two second-stage SQUID chips of the amplifier module. We acquired roughly 20 days of background events and performed the same data analysis as described previously to compare the two measurements. The resulting background spectrum is also shown in figure~\ref{fig:background} (orange, dashed). Between \SI{1}{\kilo\electronvolt} and \SI{10}{\kilo\electronvolt} we observed a background rate of \SI{1.20(8)e-4}{\per\kilo\electronvolt\per\centi\metre\squared\per\second}. The PTFE shield reduces the intensity of the copper K$_\alpha$ line by \SI{85(4)}{\percent} while the white background between \SI{1}{\kilo\electronvolt} and \SI{10}{\kilo\electronvolt} is reduced by \SI{58(3)}{\percent}. This reduction matches very well the estimation of the effectively shielded solid angle seen by the detector assuming a shield efficiency of \SI{100}{\percent} in the respective energy range.

%% file: txt/txt_conclusion.tex
\section{Conclusion}\label{sec:conclusion}

The discovery of ALPs using helioscopes requires high efficiency and low background X-ray detectors. The possibility to study the properties of ALPs implies the use of high resolution and low energy threshold detectors. Metallic magnetic calorimeters can be optimized to fulfill all these requirements. We have presented the development and characterization of the first MMC-based detector system designed to be mounted on the BabyIAXO helioscope. The detector consists of a two-dimensional 64-pixel MMC array with a fill factor of \SI{93}{\percent} covering an area of \SI{16}{\milli\metre\squared}. The absorbers of the detector are made out of \SI{20}{\micro\metre} thick gold each covering a surface of $\SI{500}{\micro\metre} \times \SI{500}{\micro\metre}$ and ensure a quantum efficiency of more than \SI{99}{\percent} for photons up to \SI{10}{\kilo\electronvolt}. A first characterization of the MMC array showed an average FWHM energy resolution of \SI{6.1}{\electronvolt} at \SI{0}{\electronvolt} and \SI{7.2}{\electronvolt} at \SI{5.9}{\kilo\electronvolt} while reaching energy thresholds below \SI{100}{\electronvolt}. The analysis of the background measured for an unshielded detector provided a background rate of \SI{3.2(1)e-4}{\per\kilo\electronvolt\per\centi\metre\squared\per\second} between \SI{1}{\kilo\electronvolt} and \SI{10}{\kilo\electronvolt}. We could attribute this background partially to fluorescence in the material surrounding the detector induced mainly by cosmic muons and radioactive impurities of our material. We have identified the possibility to reduce the background by adding a shield out of a material with a low atomic number directly above the detector. This was tested in a second characterization which showed the positive effect of the used polytetrafluoroethylene piece. The background was reduced by \SI{58(3)}{\percent} to \SI{1.20(8)e-4}{\per\kilo\electronvolt\per\centi\metre\squared\per\second} which matches the expected background reduction by the effective shielded solid angle seen by the detector. This demonstrates that a polytetrafluoroethylene shield plays already an important role to reduce the background significantly. This implies that the background can be even further reduced by the presence of active and passive shielding surrounding the detector, as already demonstrated for other detector technologies~\cite{Garza_2015}. With the results obtained in the discussed measurements we can conclude that MMCs are suitable detectors to be used in helioscopes.

%% file: txt/txt_acknowledgments.tex
\acknowledgments
We acknowledge the cleanroom team at the Kirchhoff Institute for Physics for their contribution to the fabrication of the used detector and SQUID chips. We appreciate helpful discussions and suggestions from members of the IAXO collaboration. We thank Joerg Jaeckel and Lennert Thormaehlen for providing the data for the theoretical solar axion flux. We acknowledge the screening of copper, niobium and polyimide material samples at the Canfranc underground laboratory performed by Susana Cebrián Guajardo and her team. We thank Ivor Fleck and his colleagues for the helpful discussion about the polyimide circuit boards. This work is supported by the Bundesministerium für Bildung und Forschung with the contract 05H2018-R\&D Detektoren under the project 05H18VHRD3.